\documentstyle[aps,twocolumn,epsfig,floats]{revtex}
\draft
\newcommand{\beq}{\begin{equation}}
\newcommand{\eeq}{\end{equation}}
\newcommand{\bdis}{\begin{displaymath}}
\newcommand{\edis}{\end{displaymath}}
\newcommand{\bea}{\begin{eqnarray}}
\newcommand{\eea}{\end{eqnarray}}
\newcommand{\barr}{\begin{array}}
\newcommand{\earr}{\end{array}}

\begin{document}
 
\wideabs{
\title{Universality classes in directed sandpile models}

\author{Romualdo Pastor-Satorras and Alessandro Vespignani}

\address{
The Abdus Salam International Centre for Theoretical Physics (ICTP), 
P.O. Box 586, 34100 Trieste, Italy}
 
\date{\today}
 
\maketitle
\begin{abstract}
  We perform large scale numerical simulations of a directed version
  of the two-state stochastic sandpile model. Numerical results show
  that this stochastic model defines a new universality class with
  respect to the Abelian directed sandpile. The physical origin of the
  different critical behavior has to be ascribed to the presence of
  multiple topplings in the stochastic model. These results provide
  new insights onto the long debated question of universality in
  abelian and stochastic sandpiles.
\end{abstract}
 
\pacs{PACS numbers: 05.65.+b, 05.70.Ln}
 }

The class of sandpile models, consisting of the original Bak, Tang and
Wiesenfeld (BTW) \cite{btw1btw2} automata and its theme variations, is
considered the prototypical example of a special class of driven
non-equilibrium systems exhibiting a behavior dubbed 
self-organized criticality (SOC). Under an external drive, these
systems spontaneously evolve into a stationary state. In the limit of
infinitesimal driving the stationary state shows a singular response
function associated to an avalanche-like dynamics, indicative of a
critical behavior. Sandpile models have thus attracted a great deal of
interest, as plausible candidates to explain the avalanche behavior
empirically observed in a large number of natural
phenomena\cite{jenssen98}.

In recent years, the possibility of understanding the sandpile
critical behavior in analogy with other non-equilibrium critical
phenomena such as branching processes \cite{lzs96,vz98}, interface
depinning models\cite{pacz96,la99}, and absorbing phase
transitions\cite{dvz98vdmz98,maslov94} has been pointed out.  It is
then most important to identify precisely, for sandpiles, the
universality classes and upper critical dimensions, which are basic
and discriminating features of the critical behavior.  Despite
significant numerical efforts, however, these issues remain largely
unresolved. For instance, it is still an open problem wheter or not
the original deterministic BTW sandpile and the stochastic Manna
two-state model \cite{manna91b} belong to the same universality class.
Theoretical approaches \cite{guilera92,vespignani95,hasty98} support
the idea of a single universality class, while numerical simulations
provide contradictory results \cite{grasma,milshtein98,chessa99}.

In order to have a deeper understanding of the universality classes
puzzle, we turn our attention to {\em directed sandpile models}
\cite{dhar89}.  In this case Dhar and Ramaswamy obtained an exact
solution for the Abelian directed sandpile (ADS)\cite{dhar89}, that
can be used as a benchmark to check the numerical simulation analysis.
Directed sandpiles thus become an interesting test field to study how
critical behavior is affected by the introduction of stochastic
elements.  Despite the fact that results obtained for directed models
cannot be exported ``tout court'' to the isotropic ones, the eventual
apparence of different universality classes provides interesting clues
on the general problem of universality in sandpiles.  This issue has
been recently addressed in a particular case by Tadi\'{c} and Dhar
\cite{tadic97}, but a general discussion of universality classes in
directed sandpile automata is still lacking.

In this letter we present large scale numerical simulations of Abelian
and stochastic directed sandpile (SDS) models. First, we study an ADS
model for which we recover numerically the results expected from the
analytical solution \cite{dhar89}. Then we introduce a stochastic
model which is a directed version of the Manna two-state sandpile
\cite{manna91b}. In this case, the set of critical exponents defines a
different universality class.  For both models we provide a very
accurate study of finite size effects and the convergence to the
asymptotic behavior.  For small and medium lattice sizes we find
scaling anomalies that are similar to those encountered in isotropic
models.  We also study in detail the geometrical structure of
avalanches. The presence of multiple topplings appears to be the
fundamental difference between Abelian and stochastic models.
Numerical simulations in Euclidean dimension $d>2$ show that both
universality classes have an upper critical dimension $d_c=3$, where
strong logarithmic corrections to scaling are present.

\begin{figure}[t]
  \centerline{\epsfig{file=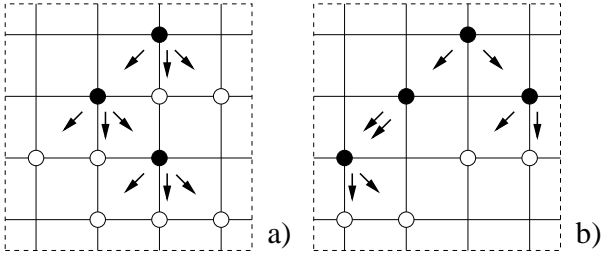, width=8cm}}
    
  \vspace*{0.5cm}
  \caption{Toppling rules in $d=2$ for directed sandpiles. Filled
    circles represent active (toppling) sites; 
    empty circles are stable sites. In the Abelian model (a) an active 
    site sends one grain to each one of its three neighbors on the
    next downwards row. In the stochastic models (b) one grain is sent
    to two randomly chosen downwards neighbors.}
  \label{fig:rules}
\end{figure}

We consider the following definition for an ADS model (see
Fig.~\ref{fig:rules}(a)): On each site of a $d$-dimensional hypercubic
lattice of size $L$, we assign an integer variable $z_i$, called
``energy''. Each time step, an energy grain is added to a randomly
chosen site ($z_i \to z_i + 1$). When a site acquires an energy greater
than or equal to the threshold $z_c=2d-1$, it topples. Topplings are
directed along a fixed direction $x_\parallel$ (defined usually as
``downwards''): when a site on the hyperplane $x_\parallel$ topples, it sends
{\em deterministically} one energy grain to each nearest and
next-nearest neighbor site on the hyperplane $x_\parallel+1$, for a total of
$2d-1$ grains.  This definition differs from the ADS studied by Dhar
and Ramaswamy \cite{dhar89} in the orientation of the lattice.  Both
models, however, share the same universality class, being Abelian,
deterministic, and directed.

The stochastic generalization of the above model is depicted in
Fig.~\ref{fig:rules}(b): The threshold is now $z_c=2$, independent of
the spatial dimension. When a site at the hyperplane $x_\parallel$ topples, it
sends two grains of energy to two sites, randomly chosen among its
$2d-1$ neighbors in the hyperplane $x_\parallel+1$.  The dynamical rule of
this model can be defined {\em exclusive} if the two energy grains are
always distributed on different sites.  On the contrary, a {\em
  nonexclusive} dynamics allows the transfer of two energy grains to
the same site. We will consider separately the cases of the exclusive
stochastic directed sandpile (ESDS) and the nonexclusive stochastic
directed sandpile (NESDS).  It is worth remarking that stochasticity
does not alter the Abelian nature of the model \cite{dhar99}.  All
three models are locally conservative; no energy grains are lost
during a toppling event.  Boundary conditions are periodic in the
transverse directions and open at the bottom hyperplane $x_\parallel=L$, from
which energy can leave the system.

In the critical stationary state, we can define the probability that
the addition of a single grain is followed by an avalanche of toppling
events. Avalanches are then characterized by the number of topplings
$s$, and the duration $t$. According to the standard finite size 
scaling (FSS) hypotesis, the probability distributions of these
quantities are described by the scaling functions \beq
P(s)=s^{-\tau_s}{\cal G}(s/s_c)\ ,
\label{ps}
\eeq
\beq 
P(t)=t^{-\tau_t}{\cal F}(t/t_c)\ ,
\label{pt}
\eeq where $s_c$ and $t_c$ are the cutoff characteristic size and
time, respectively. In the critical state the lattice size $L$ is the
only characteristic length present in the system. Approaching the
thermodynamic limit ($L\to\infty$), the characteristic avalanche size and
time diverge as $s_c\sim L^D$ and $t_c\sim L^z$, respectively. The exponent
$D$ defines the fractal dimension of the avalanche cluster and $z$ is
the usual dynamic critical exponent. The directed nature of the model
introduces a drastic simplification, since it imposes $z=1$. A general
result concerns the average avalanche size $\left<s\right>$, that also
scales linearly with $L$\cite{dhar89,kad89,tsuchiya99}: a new injected
grain of energy has to travel, on average, a distance of order $L$
before reaching the boundary. In the stationary state, to each energy
grain input must correspond, on average, an energy grain flowing out
of the system.  This implies that the average avalanche size
corresponds to the number of topplings needed for a grain to reach the
boundary; i.e.  $\left<s\right>\sim L$.  The same result can be exactly
obtained by inspecting the conservation symmetry of the model
\cite{pv99}.

For the ADS, the exact analytical solution in $d=2$ yields the
exponents $\tau_s=4/3$, $\tau_t=3/2$ and $D=3/2$ \cite{dhar89}.  The upper
critical dimension is found to be $d_c=3$, and it is also possible to
find exactly the logarithmic corrections to scaling
\cite{dhar89,lubeck98}.  The introduction of stochastic ingredients in
the toppling dynamics of directed sandpiles has been studied only
recently in a model that randomly stores energy on each toppling
\cite{tadic97}. This model is strictly related to directed percolation
and defines a universality class ``per se''.  In our case
stochasticity affects only the partition of energy during
topplings, and there is no analytical insight for the critical
behavior of this model. In order to discriminate between ADS and SDS we
perform simulations of both models for sizes ranging from
$L=100$ to $L=6400$.  Statistical distributions are obtained averaging
over $10^7$ avalanches.  Comparison of numerical results on the ADS
allows us to check the reliability and degree of convergence with
respect to the lattice sizes used.

It is well known from the many numerical papers on sandpiles that an
accurate determination of the exponents $\tau_s$ and $\tau_t$ is a subtle
issue.  An overall determination within a $10\%$ of accuracy is a
relatively easy task.  However, a truly accurate measurement, allowing
a precise discrimination of universality classes, is strongly affected
by the lower and upper cut-offs in the distribution.  Extrapolations
and local slope analysis are often very complicated and the relative
error bars are not clearly defined.  In this respect, it is far better
to calculate exponents by methods that contain the system-size
dependence explicitly; namely data collapse and moment analysis.
Moment analysis was introduced by De Menech {\em et al.}
\cite{mentebaldi99} in the context of the two dimensional BTW, and it
has been used extensively on Abelian and stochastic models
\cite{chessa99,granada}.  The $q$-moment of the avalanche size
distribution on a lattice of size $L$, $\left\langle s^q\right\rangle_L=\int s^q
P(s)ds$, has the following size dependence
\begin{equation}
\langle s^q\rangle_L= L^{D(q+1-\tau_s)}\int y^{q-\tau_s}{\cal G}(y)dy
\sim L^{D(q+1-\tau_s)},
\label{eq:sq}
\end{equation}
where we have used the transformation $y=s/L^{D}$ in the
finite size scaling (FSS) form Eq.~(\ref{ps}).  More generally,
$\left\langle s^q\right\rangle_L\sim L^{\sigma_s(q)}$, where the exponents $\sigma_s(q)$ can be
obtained as the slope of the log-log plot of $\left<s^q\right>_L$
versus $L$.  Using Eq.~(\ref{eq:sq}), we obtain $\langle s^{q+1}\rangle_L/\langle s^q\rangle_L
\sim L^{D}$ or $\sigma_s(q+1)-\sigma_s(q)=D$, so that the slope of $\sigma_s(q)$ as a
function of $q$ is the cutoff exponent; i.e.  $D=\partial\sigma_s(q)/\partial q$.  This
is not true for small $q$ because the integral in
Eq.~(\ref{eq:sq}) is dominated by its lower cutoff.  In particular,
corrections to scaling are important for $q \leq\tau_s-1$.  An additional
and strong check on the numerical data can be found
in the fact that, as we have previously shown, the first moment of the
size distribution must scale linearly with $L$.  This last constrain
also allows the evaluation of the the exponent $\tau_s$ from the scaling
relation $(2-\tau_s)D=\sigma_s(1)=1$, that should be satisfied for large
enough sizes.

\begin{figure}[t]
  \centerline{\epsfig{file=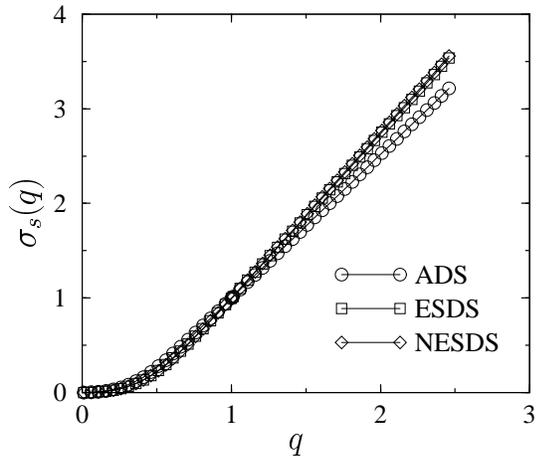, width=7cm}}
  \caption{Plot of $\sigma_s(q)$ for the $d=2$ models
    ADS, ESDS, and NESDS.}
    
  \label{fig:moments}
\end{figure}

\begin{table}[b]
\begin{tabular}{ccccc}

Model & $\tau_s$           & $D$           & $\tau_t$           & $z$ \\
\hline
DR   & $4/3$       & $3/2$       & $3/2$       & $1$\\ 
ADS  & $1.34\pm0.01$ & $1.51\pm0.01$ & $1.51\pm0.02$ & $1.00\pm0.01$\\ 
ESDS & $1.43\pm0.01$ & $1.74\pm0.01$ & $1.71\pm0.03$ & $0.99\pm0.01$\\ 
NESDS& $1.43\pm0.01$ & $1.75\pm0.01$ & $1.74\pm0.04$ & $0.99\pm0.01$\\ 
\end{tabular}
\caption{Critical exponents for directed sandpiles in $d=2$. DR: Dhar 
  and Ramaswamy's exact result; ADS: Abelian model; ESDS, NEDSD:
  stochastic models.}
\label{table}
\end{table}

Along the same lines we can obtain the moments of the avalanche time
distribution. In this case $\langle t^q\rangle_L\sim L^{\sigma_t(q)}$, with $\partial\sigma_t(q)/\partial
q=z$. Analogous considerations for small $q$ apply also for the time
moment analysis.  Here, an estimate of the asymptotic convergence of
the numerical results is provided by the constraint $z=1$, that 
holds for large enough sizes. Then, the $\tau_t$ exponent can be found
using the scaling relation $(2-\tau_t)=\sigma_t(1)$.

\begin{figure}[t]
  \centerline{\epsfig{file=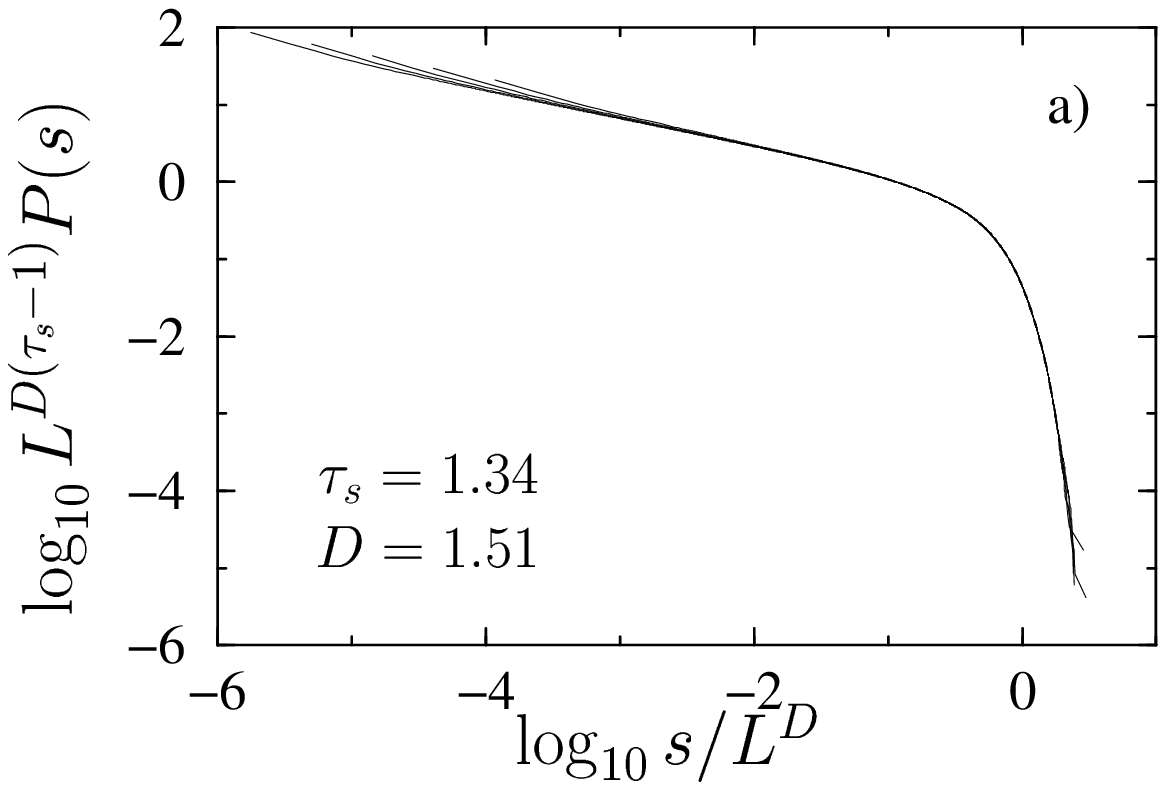, width=7cm}}
  \vspace*{0.2cm}
  \centerline{\epsfig{file=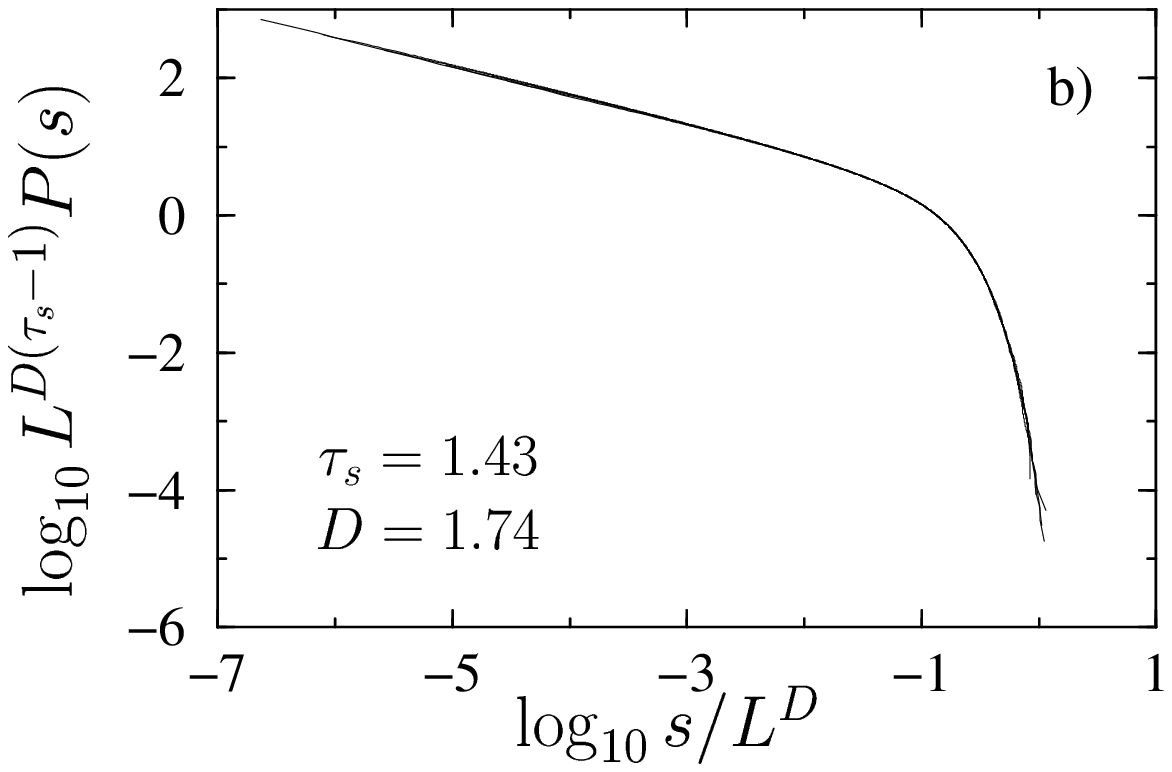, width=7cm}}
  \vspace*{0.2cm}

 \caption{Data collapse analysis of the integrated avalanche size
   distribution for the $d=2$ models a) ADS and b) ESDS.  System sizes
   are $L=400, 800, 1600, 3200$, and $6400$. The results for the ESDS
   and NESDS are identical, within error bars.}
  \label{fig:sizes}
\end{figure}

\begin{figure}[t]
  \centerline{\epsfig{file=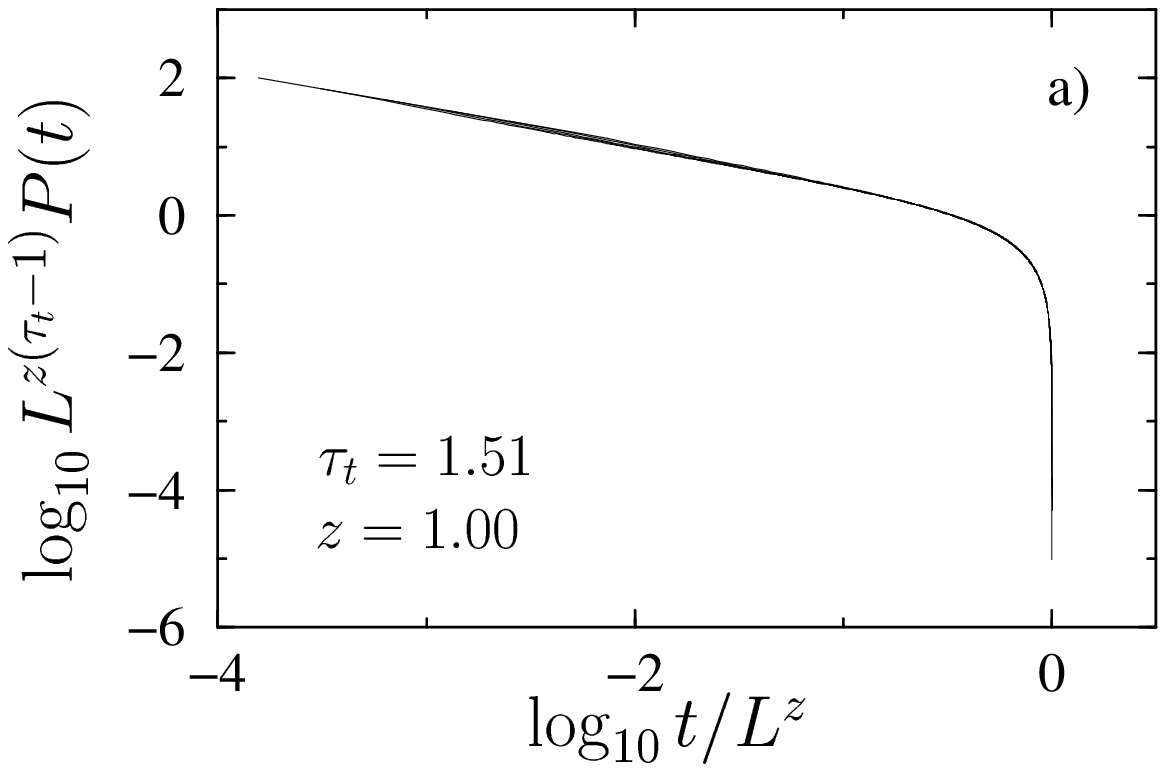, width=7cm}}
  \vspace*{0.2cm}
  \centerline{\epsfig{file=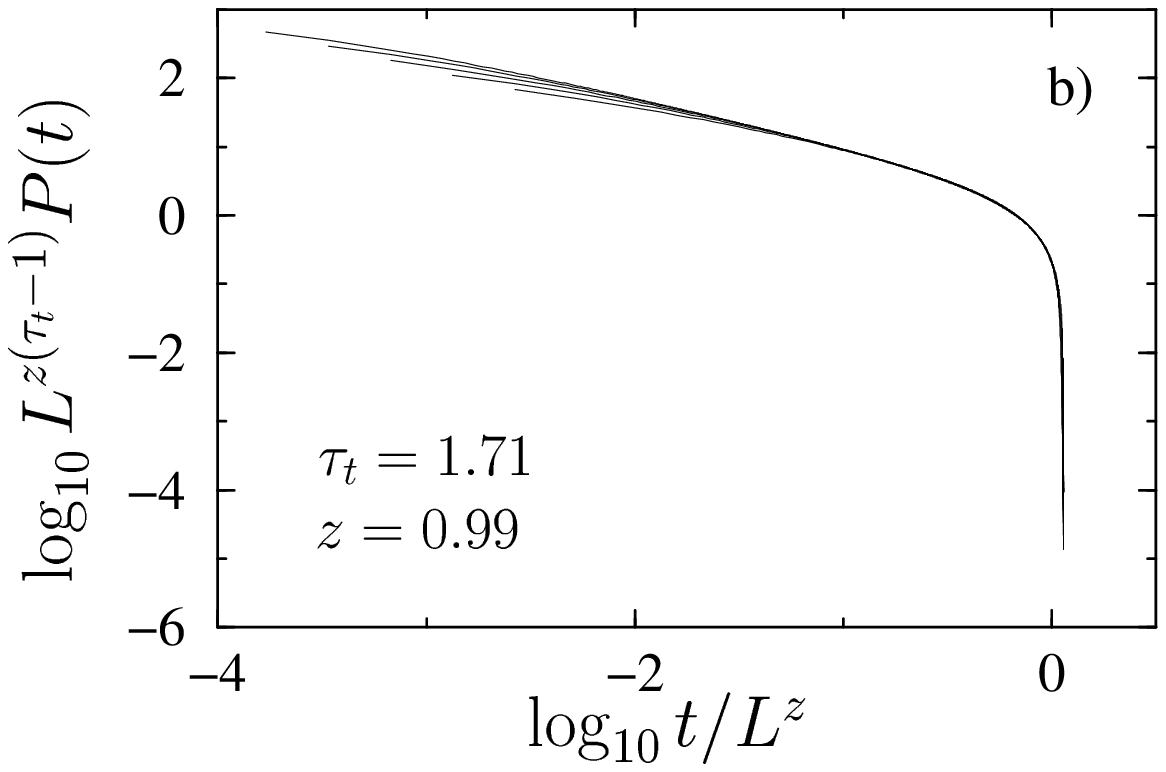, width=7cm}}
  \vspace*{0.2cm}

  \caption{Data collapse analysis of the integrated avalanche time
    distribution for the $d=2$ models a) ADS and b) ESDS.  System
    sizes are $L=400, 800, 1600, 3200$, and $6400$. The results for
    the ESDS and NESDS are identical, within error bars.}
  \label{fig:times}
\end{figure}

Despite the fact that the moment method is usually rather accurate,
it must be corroborated by a data collapse analysis. The FSS of
Eqs.~(\ref{ps},\ref{pt}) has to be verified and must be consistent
with the numerical exponents obtained from the moment analysis. This
can be done by rescaling $s\to s/L^D$ and $P(s)\to P(s)L^{D\tau_s}$ and
correspondingly $t\to t/L^z$ and $P(t)\to P(t)L^{z\tau_t}$. Data for
different $L$ must then collapse onto the same universal curve if the
FSS hypothesis is satisfied. Complete consistency between the methods
gives the best collapse with the exponents obtained by the moments
analysis.  In Table~\ref{table} we report the exponents found for the
ADS, ESDS and NESDS in $d=2$.  Figure~\ref{fig:moments} shows the
moments $\sigma_s(q)$. Figures~\ref{fig:sizes} and \ref{fig:times} plot the
FSS data collapse for sizes and times, respectively.

\begin{figure}[t]
  \centerline{\epsfig{file=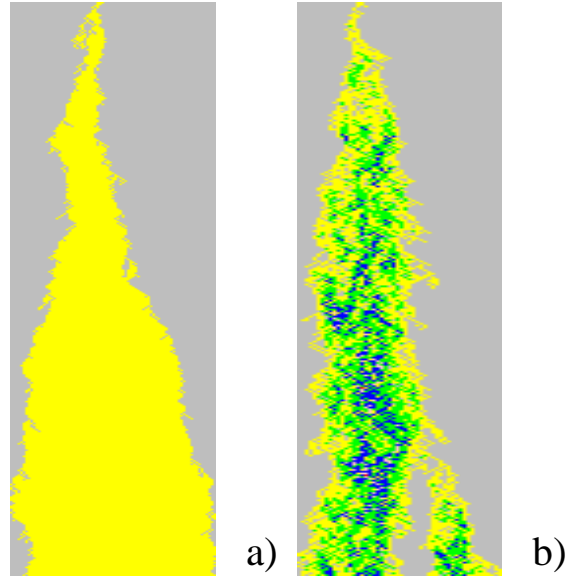, width=7.5cm}}
  \vspace*{0.1cm}
  \caption{Color plots of the local density of topplings in
    two avalanches of size $50 000$, in $d=2$, for a) ADS and b) ESDS.
    Yellow represents a single toppling per site; red stands for the
    maximum number of topplings.}
  \label{fig:profiles}
\end{figure}

The exponents obtained for the ADS are in perfect agreement with the
expected analytical results. This fact supports the idea that the
system sizes used in the present work allow to recover the correct
asymptotic behavior. It is worth remarking that, for small and medium
lattice sizes, both moments and data collapse analysis present scaling
features that can not be reconciled in the single scaling picture
usually considered. These anomalies are not persistent and disappear
for reasonably large sizes ($L\simeq 10^3$). This evidence for a slow
decaying of finite size effects could shed light into several
anomalies reported in isotropic sandpiles, for which, unfortunately,
it is very difficult to reach very large sizes
\cite{chessa99,mentebaldi99,granada}.  Results for the ESDS and NESDS
are identical within the error bars, indicating that these two
models are in the same universality class. On the other hand, the
obtained exponents show, beyond any doubt, that Abelian and stochastic
directed sandpile models do not belong to the same universality class.

\begin{figure}[t]
  \centerline{\epsfig{file=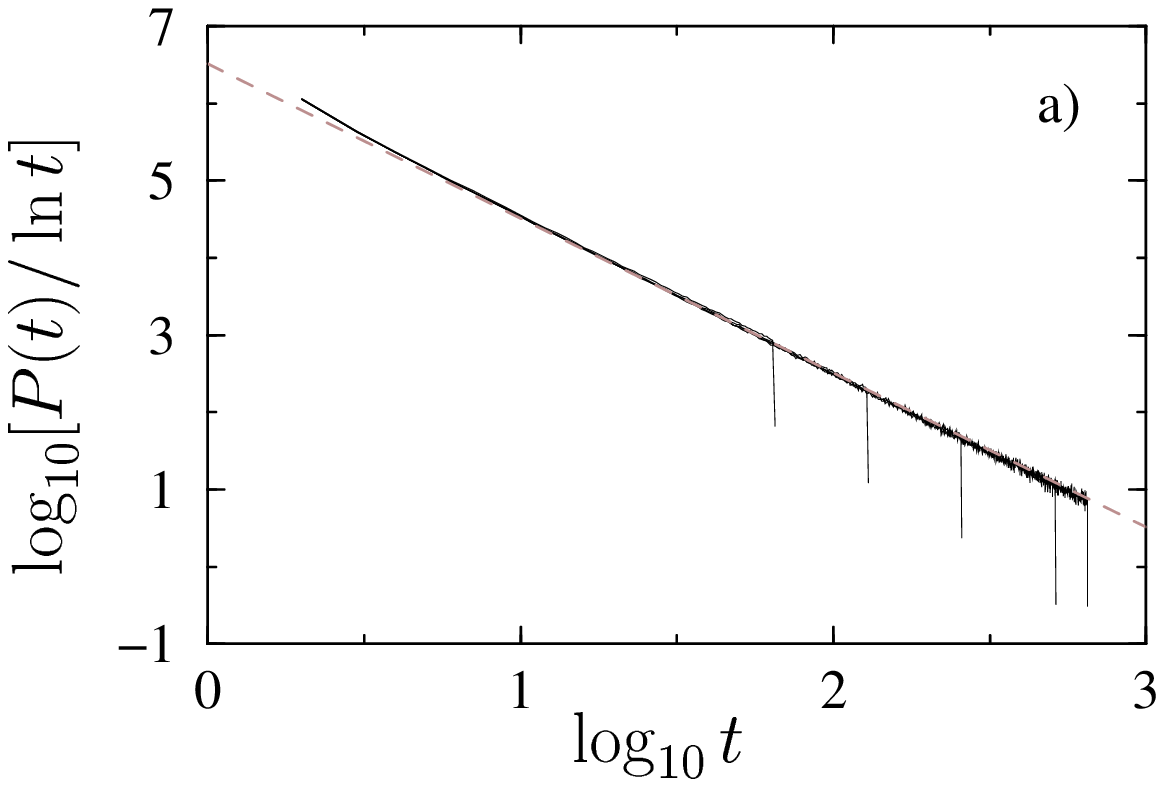, width=7cm}}
  \vspace*{0.2cm}
  \centerline{\epsfig{file=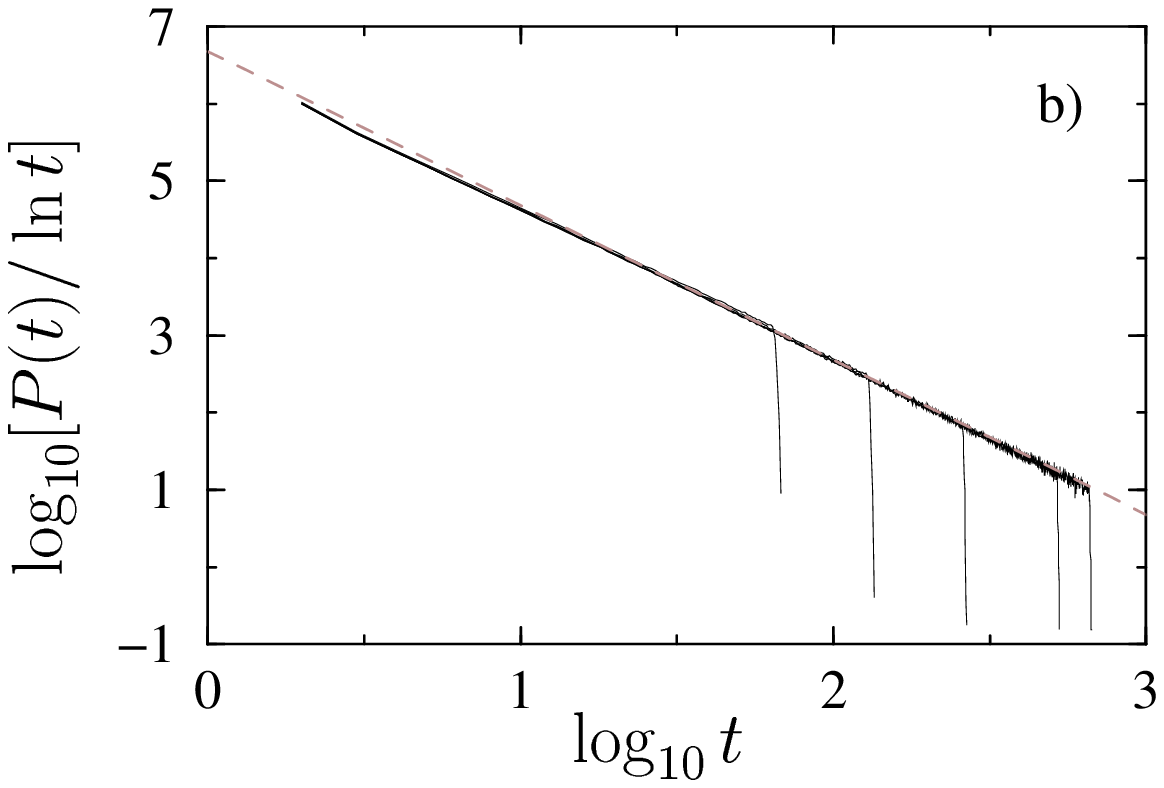, width=7cm}}
  \caption{Data collapse with logarithmic corrections of the avalanche 
    time distribution for the $d=3$ models a) ADS and b) ESDS. The
    dashed lines have slope $-2$. System sizes are $L=64, 128, 256,
    512$, and $650$.}
  \label{fig:3dtimes}
\end{figure}

The compelling numerical evidence for two distinct universality
classes does not tell us what is the basic mechanism at the origin of
the different critical behavior. In order to have a deeper insight
into the dynamics of the various models, we have inspected the
geometric structure of the resulting avalanches.  In
Fig.~\ref{fig:profiles} we depict in a color plot the local density
of topplings in two avalanches of size 50 000 corresponding to the
two-dimensional ADS and ESDS models. From the figure it becomes
apparent that the stochastic dynamics introduces multiple toppling
events, which are by definition absent in the Abelian case. This gives
rise to very different avalanche structures, eventually reflected in
the asymptotic critical behavior. In particular, the fractal dimension
$D$ is indicative of the scaling of toppling events with sizes. In the
stochastic case we recover a higher fractal dimension than in the
Abelian case. The multiple toppling mechanism has been proposed in the
past as the origin of differences between {\em isotropic} Abelian and
stochastic sandpiles as well.  In that case, however, multiple
toppling is a common feature of both models, and for the largest sizes
reached so far share, they share the same fractal dimension $D$
\cite{chessa99}.

Analysis of the models in three dimensions is strongly hindered by the
presence of logarithmic corrections \cite{dhar89,lubeck98}.
Nonetheless, a naive application of the moment analysis yields values
compatible with the mean-field results $\tau_s=3/2$, $\tau_t=2$, and $D=2$
\cite{dhar89}. More interestingly, in Ref.~\cite{dhar89} the
authors were able to deduce the exact form of the logarithmic
corrections in $d=3$ for the avalanche time distribution, namely $P(t)\sim
t^{-2} \ln t$. In Figure~\ref{fig:3dtimes} we have checked that the
same logarithmic corrections apply to both the Abelian and
ESDS sandpiles. This remarkable fact lends support to
the critical dimension of the stochastic model being $d_c=3$.

In summary, we have reported large scale numerical simulations of a
stochastic directed sandpile model. This model defines unambigously a
different universality class with respect to the Abelian directed
sandpile model. The origin of this difference is traced back to the
avalanche cluster geometric structure, providing new clues to
understand the effect of stochastic elements in the dynamics of
avalanche processes.

\vspace*{0.5cm}

This work has been supported by the European Network under Contract
No.  ERBFMRXCT980183. We thank D. Dhar, D. Dickman, M. A. Mu{\~n}oz, A.
Stella, and S. Zapperi for helpful comments and discussions.

%\bibliographystyle{prsty}
%\bibliography{soc}

\end{document}